\documentclass{nature}
\usepackage{graphicx}
\usepackage{amssymb}
\usepackage{amsfonts}
\usepackage{caption}
\usepackage{bm}
\usepackage{epstopdf}
\usepackage{epsfig}
\usepackage{color}
\usepackage{amssymb, amsmath}
\begin{document}

\title{Mott transition and Antiferromagnetic Metal on Shastry-Sutherland Lattice}

\author{Hai-Di Liu, Yao-Hua Chen, Heng-Fu Lin, Hong-Shuai Tao, Wu-Ming Liu$^{\star}$}

\maketitle

\begin{affiliations}
\item Beijing National Laboratory for Condensed Matter Physics, Institute of Physics, Chinese Academy of Sciences, Beijing 100190,
China.

$^\star$e-mail: wliu@iphy.ac.cn
\end{affiliations}

\begin{abstract}
The Shastry-Sutherland lattice, one of the simplest systems with geometric frustration, which has an exact eigenstate by putting singlets on diagonal bonds, can be realized in a group of layered compounds and rises both theoretical and experimental 
interest. 
Most of the previous studies on the Shastry-Sutherland lattice are focusing on the Heisenberg model. Here we opt for the Hubbard model to calculate phase diagrams over a wide  range of interaction parameters, and show the competing effects of interaction, frustration and temperature. 
At low temperature, frustration is shown to favor a paramagnetic metallic ground state, while interaction drives the system to an antiferromagnetic insulator phase. 
Between these two phases, there are an antiferromagnetic metal phase and a paramagnetic insulator (which should be a valence bond solid) phase  resulting from the competition of the frustration and the interaction. Our results may shed light on more exhaustive studies about quantum phase transitions in this lattice.

\end{abstract}
  Strongly correlated system with geometric frustration is an important research area in condensed matter physics, and has attracted enormous interests in recent years.
  Geometric frustration refers to the case where the geometry of a lattice conflicts with its inter-site interactions. It can be realized in antiferromagnetic Heisenberg models on certain lattices such as triangular lattice \cite{Parcollet} and kagome lattice  \cite{Ohashi} which host competing exchange interactions that can not be satisfied simultaneously. This can give rise to new possibilities for the ground state, including the usual antiferromagnetic order and many novel states, such as heavy Fermi liquid \cite{HF1,HF2,HF3}, spin liquid\cite{SL1,SL2,SL3,SL4}, spin glass\cite{SG1,SG2,SG3,SG4}, and spin ice\cite{SI1,SI2,SI3,SI4}, which are poorly understood and thus actively studied.

 The Shastry-Sutherland lattice is one example of geometric frustrated systems, and has been actively researched due to the magnetization plateaus in the presence of a magnetic field\cite{PhysRevLett.87.097203,PhysRevLett.101.177201,PhysRevLett.101.227201,PhysRevLett.109.167202,PhysRevE.88.022111}. It was first proposed by Shastry and Sutherland \cite{SriramShastry19811069}  as a theoretical toy model with the Heisenberg Hamiltonian where there are exchange interactions on the nearest bond as well as the diagonal bonds (see in Fig.~1 (b)). It is shown to have an exact eigenstate consist of orthogonal singlet dimers on the diagonal bonds, which becomes the exact ground state of the system when frustration is strong.   
 It is later found out that the Shastry-Sutherland lattice can be used to describe the magnetic properties of the compound $SrCu_2(BO_3)_2$ \cite{Smith1991430,PhysRevLett.82.3168,0953-8984-15-9-201}, in which the two dimensional magnetic linkage of the $Cu^{2+}$ ions has a structure  showed in Fig.~1 (a) and is topologically equivalent to Shastry-Sutherland lattice. Similar structures have also been found in other materials such as $Yb_2Pt_2Pb$ \cite{PhysRevLett.110.017201} and $RB_4$($R=La-Lu$) \cite{PhysRevE.88.022111,PhysRevLett.109.167202}.
 Most of the theoretical and numerical studies on Shastry-Sutherland lattice so far have been focusing on Heisenberg model \cite{PhysRevLett.82.3701,PhysRevLett.84.4461,PhysRevB.87.144419},  Ising model \cite{PhysRevLett.109.167202,PhysRevE.88.022111}, and $t-J$ model\cite{PhysRevLett.99.227003,PhysRevLett.93.207004}. In this report, we opt for the Hubbard model to investigate correlated electrons on this geometrical frustrated system, which degenerates to the Heisenberg model in the large U limit at half-filling.
 Our study based on Hubbard model thus can provide phase diagrams that include a wider range of interaction parameters.

To understand this strong correlated many-body system with geometrial frustration, we apply the cellular dynamical mean field theory (CDMFT) combined with the continuous time quantum Monte Carlo method (CTQMC). 
The CDMFT, which incorporates the short-range spatial correlations by mapping the lattice to a self-consistent embedded cluster in real space instead of a single site in dynamical mean field theory (DMFT)\cite{RevModPhys.68.13}, has been proved to be successful when applied to study strong interaction systems with geometric frustration\cite{PhysRevLett.87.186401,PhysRevLett.101.186403,RevModPhys.77.1027}. 
The CTQMC, on the other hand, which is more accurate than the traditional quantum Monte Carlo method, is used as an inpurity solver \cite{MC, MCRMP}.  From the single-particle Green's function given by the CDMFT and CTQMC, the single-particle density of states and the double occupancy can be calculated, which are further used to identify the Mott metal-insulator transition. Due to the presence of frustration, magnetic order is another important aspect that we would like to address in this report, which can also be extracted from the single-particle Green's function. Then we obtain  phase diagrams including the effect of interaction, frustration and temperature. 
At low temperature, apart from the antiferromagnetic insulator and paramagnetic metal phase that usually appear in a Mott transition, the competition between  frustration and interaction gives rise to two other phases. One is an antiferromagnetic metal phase in the intermediate interaction region before the onset of the metal-insulator transition. The other one is a paramagnetic insulator phase at both large interactions and large frustrations. We speculate it to be a valence bond solid state, where pairs of spins form spin-0 singlets \cite{Balents}.

\section*{Results}
The standard Hubbard model on the Shastry-Sutherland lattice can be written as
\begin{equation}
H=-t_{1}\sum\limits_{\langle ij \rangle_{1},\sigma}c_{i\sigma }^{+}c_{j\sigma}-t_{2}\sum\limits_{\langle ij \rangle_{2},\sigma}c_{i\sigma }^{+}c_{j\sigma}+U\sum\limits_{i}n_{i\uparrow}n_{i\downarrow },
\label{H}
\end{equation}
where $c_{i\sigma}^{+}$ ($c_{i\sigma}$) is the creation (annihilation) operator; $n_{i}=c_{i}^{+}c_{i}$ is the number operator; $t_{1}$ ($t_{2}$) is the nearest-neighbor (the diagonal) hopping energy as shown in Fig.~1(b). $U$ is the Hubbard interaction strength. $\langle ij\rangle _{1}$ ($\langle ij \rangle_{2}$) runs over the nearest-neighbor links (the diagonal links) on the lattice. We set $t_{1}=1.0$ as the energy unit, and $t_{2}$ hence can be seen as a measure of the frustration strength. We focus on the half-filling case i.e. $ \langle n \rangle=1 $, which is realized by adjusting the chemical potential $\mu$.

We shall start from the non-interacting case with $U=0$. In this case, the Hamiltonian in Eq.~\eqref{H} can be partially diagonalized in momentum space as $H_{0}=\sum\limits_{k}\psi _{k}^{+}t(k)\psi _{k}$, where $\psi _{k}^{+}=(c_{1k}^{+},c_{2k}^{+},c_{3k}^{+},c_{4k}^{+})$. The index $i =1,2,3,4$ in $c_{ik}$ represent the four sites in each unit cell as illustrated in Fig.~1 (b), and $k$ locates in the first Brillouin zone. $t(k)$ is a $4\times4$ matrix that has a form,
\begin{eqnarray*}
t(k)&= %
&\begin{bmatrix}
0 & -2t_{1}\cos k_{x} & -t_{2}e^{-ik_{x}-ik_{y}} & -2t_{1}\cos k_{y}\\
-2t_{1}\cos k_{x} & 0 & -2t_{1}\cos k_{y} & -t_{2}e^{-ik_{x}+ik_{y}}\\
-t_{2}e^{ik_{x}+ik_{y}} & -2t_{1}\cos k_{y} & 0  & -2t_{1}\cos k_{x}\\
-2t_{1}\cos k_{y} & -t_{2}e^{ik_{x}-ik_{y}} & -2t_{1}\cos k_{x} & 0  %
\end{bmatrix}.%
\end{eqnarray*}

 The band dispersion is readily obtained by diagonalizing this matrix. A special case with $t_2=1.0$ is shown in Fig.~1 (d), where there are four bands resulting from the four sub-lattices. We label them I II III and IV from top to bottom, among which band II and III touch at $\Gamma$ point ($k=(0,0)$). Band II has its minimum at $\Gamma$ and two diagonals in the first Brillouin zone, which result in a Von Hove singularity right at half filling in the density of states as shown in Fig.~1(e). 
 In addition, we show in Fig.~2 the evolution of the band dispersions along the high symmetry lines of the first Brillouin zone with respect to $t_2$. It can be seen that there are several things in common for the dispersions at different $t_2$'s. For example, the band is a flat along the diagonal of the first Brillouin zone. Along the boundary of the first Brillouin zone, band I and II as well as band III and IV are degenerate. 
 Besides, when $t_2>2.0$, there is a gap opened between band II and III, and the system becomes a band insulator at half-filling. As we are focusing on the Mott transition, we limit our studies to the case when $t_2<2.0$ in this report.

Now we turn to the case when the Hubbard interaction is represent (i.e. $U>0$). A large U can induce a large energy cost when two electrons occupy the same site, leading to a decreasing double occupancy. 
Consequently, at half filling electrons tends to get localized with one electron per site and when $U$ is large enough the system becomes an insulator with gap in the single-particle electron spectrum, known as the Mott metal-insulator transition. 
In order to identify the Mott transition of the Shastry-Sutherland lattice, we calculated the density of states (DOS) at different interaction strength by the maximum entropy method \cite{Jarrell1996133} with different temperature and frustration.  
In Fig.~3(a), we show the DOS for $T=0.2,t_2=1.0$. It can be seen that when $U=7$ a Fermi-liquid-like peak is found near Fermi energy, which splits to a pseudogap when $U$ is increased to $8.5$\cite{chenyaohuaL,chenyaohuaA,PhysRevLett.86.139,PhysRevB.65.233103}. When $U$ increases up to the critical point $U=9.5$ an obvious gap appears around the Fermi level which suggests the system undergoes a Mott transition from a metal to a Mott insulator. In Fig.~3(b) we plot the DOS at $T=0.1$ with $t_2=1.0$ and we can see that  the transition point at $T=0.1$ decreases to $U=8$ compared with $U=9.5$ when $T=0.2$, which is because the decreasing of temperature suppresses the thermal fluctuations and makes the Mott transition easier. 
By comparing Figs.~3(b)-3(d), it is found out that the critical interaction of Mott transition increases when the frustration becomes stronger.

The double occupancy $Docc=\langle n_{i\uparrow}n_{i\downarrow} \rangle $ represents the probability of two particles occupying  the same site. The evolution of $Docc$ as a function of $U$ at different temperature is shown in Fig.~4. 
It can be seen that as $U$ increases,  $Docc$ decreases monotonically to a small value, which is a characteristic feature of the Mott transition \cite{wuwei,chenyaohuaL,chenyaohuaA,Ohashi}, indicating that  the system is almost singly occupied at large $U$. 
 Meanwhile, $Docc$ has no visible discontinuities at the critical points marked with arrows, suggesting that the Mott transition here is continuous. We also observe a natural decrease of the critical $U$ when the temperature decreases, which suppresses the thermal fluctuations.
Besides, the evolution of Docc as a function of the frustration strength for different $U$ is shown in the inset of Fig.~4, and it can be seen that the double occupancy is increased with increasing $t_2$ or decreasing $U$.

There is usually magnetic order developed along with the Mott metal-insulator transition.
In order to investigate the formation of the magnetic order, we define a staggered magnetic order parameter as $ m=\frac{1}{N}\sum_{i}sign(i)(\langle n_{i\uparrow } \rangle - \langle n_{i\downarrow } \rangle) $, where $sign(i)=1$ if $i=1,3$ and $sign(i)=-1$ if $i=2,4$ as shown in Fig.~1(b). 
Fig.~5 shows the evolution of this staggered magnetic order parameter $m$ and the single-particle gap $\triangle E$ as a function of $U$ for $T=0.1$, with $t_2=1.0$. When $U=6$ both the staggered magnetic order parameter and the single-particle gap vanish when, and the system is in a paramagnetic metal phase. 
When $6<U<8$, the magnetic order forms while the single-particle excitation is still gapless, implying the system is in an antiferromagnetic metal phase. 
When $U$ increases to $8$, a gap opens, and the system goese into an antiferromagnetic insulator phase. 
In the inset of Fig.~5, we also show the case when $t_2=1.4$. There only exists the metal-insulator transition at $U=10$, and all magnetic orders are suppressed due to strong frustration.

Carrying out the analysis above in a wide range of parameters, phase diagrams of interacting fermions on the Shastry-Sutherland lattice with respect to interaction, frustration, and temperature are obtained in Figs.~6 and 7. 
We first show the effect of frustration $t_2$  and interaction $U$ at a low enough temperature $T=0.1$ (see Fig.~6). 
When $t_2=0$, the Shastry-Sutherland Lattice degenerates to a square lattice, which is unstable towards an antiferromagnetic phase for an arbitrarily small interaction due to the perfect nesting of the Fermi surface, and the magnetic order is always accompanied by a Mott insulating gap\cite{PhysRevLett.87.167010,RevModPhys.77.1027}.
In Fig.~6, the finite critical $U$ which is around 2.9 when $t_2=0$ is due to the finite temperature. 
In the presence of frustration, when $U$ increases, instead of going directly into the antiferromagnetic insulator phase from the paramagnetic metal phase, the system first enters an antiferromagnetic metal phase, showing an important role that frustration plays in the formation of an  antiferromagnetic metal
\cite{PhysRevLett.83.2386, PhysRevB.55.R676, PhysRevB.49.10181}.  
When $t_2>1.3$ the antiferromagnetic metal phase disappear and a paramagnetic insulator phase emerges for $U>9.5$.
The critical frustration $t_2 = 1.3$ obtained here is consistent with the one obtained from the calculations of the two dimensional Heisenberg  model on the Shastry-Sutherland lattice, beyond which the ground state of the system was shown to become a valence bond solid\cite{PhysRevLett.82.3701,0953-8984-15-9-201,PhysRevB.60.6608,PhysRevB.72.094436,PhysRevE.74.026701}. We will return to this later in more details in the discussion part.

We also obtain phase diagrams showing the effects of thermal fluctuations at a fixed frustration strength $t_2=1.0$, as illustrated in Fig.~7.  
The antiferromagnetic metal phase exists when $T<0.175$, and a phase transition from antiferromagnetic metal to antiferromagnetic insulator is found when $U$ is kept increasing.
When temperature is high, magnetic orders are suppressed by thermal fluctuations, and the system undergoes a transition from a paramagnetic metal phase to a paramagnetic insulator phase with increasing interaction. 
Additionally, we plot the phase digram with frustration up to $1.3$ in the inset of Fig.~7. 
It is shown that due to the strong frustration, magnetic  orders are totally suppressed even at low temperature;
both the antiferromagnetic metal phase and the antiferromagnetic insulator phase disappears. 
When the interaction is increased, there is only a transition from a paramagnetic metal to a paramagnetic insulator. 

Finally, we present  in Fig.~8 the distribution of the spectral weight at zero frequency, $A(k,\omega\!=\!0) \approx -\frac{1}{\pi} \lim {\omega_{n \rightarrow 0}}ImG(k, i\omega_{n})$ for different $U$ and $t_2$ with $T=0.1$. 
The location of the maxima of $A(k,\omega\!=\!0)$ can be seen as the Fermi surface \cite{Parcollet}. 
As showed in Fig.~8, when the interaction is small, the spectral function has sharp peaks at the center and along the two intersecting diagonals  of the first Brillouin zone, which is weakly renormalized compared to the non-interaction case, and exhibits a well-defined Fermi surface.
With the increasing of the interaction, the peaks become lower, and finally vanish when the Mott transition happens due to the localization of particles. The decreasing of the frustration also makes the Fermi surface shrink.

\section*{Discussion}
As mentioned in the introduction part, most of the theoretical and numerical studies of Shastry-Sutherland lattice are focusing on localized spins based on the Heisenberg model \cite{PhysRevLett.82.3701,PhysRevLett.84.4461,PhysRevB.87.144419}, whose ground state phase diagram has two limiting behavior depending on the dimensionless parameter $J_1/J_2$, where $J_1$ is the exchange coupling constant along the nearest neighbour bonds and $J_2$ the one along the additional diagonal bonds. In the limit $J_1/J_2\gg1$, the ground state is in antiferromagnet with gapless magnetic excitations. In the opposite limit $J_1/J_2\ll1$, the exact ground state is proved to be a valence bond solid where local spins form singlet dimers (see Fig.~6 (a3)) on the diagonal bonds. The transition point between these two phases at $T=0$ has been estimated to be $J_1/J_2\approx0.6 \sim 0.7$ \cite{PhysRevLett.82.3701,0953-8984-15-9-201,PhysRevB.60.6608,PhysRevB.72.094436,PhysRevE.74.026701}.

In this report, we numerically investigate the Hubbard model on the Shastry-Sutherland lattice which takes the itinerancy of electrons into account.
When $U$ is large and overweights the effect of the kinetic energy, electrons get localized and the system reduces to the antiferromagnetic Heisenberg model, which serves as a benchmark of our calculations. As shown in Fig.~6, at large $U$ limit, spins order antiferromagnetically when frustration ($t_2/t_1$) is small. 
When frustration goes large, magnetic orders are suppressed and the system is a paramagnetic insulator. The critical point is around $t_2/t_1=1.3$, which, at large $U$ limit, corresponds to $J_1/J_2=(t_1/t_2)^2 \approx 0.6$ for a Heisenberg model. This is in good agreement with  previous predictions, as mentioned in References [26]-[30], where the system has a valence bond solid as its exact ground state beyond a critical $J_1/J_2 \approx 0.6 \sim 0.7$. Therefore, the paramagnetic insulator state that appears when $t_2>1.3$ should be a valence bond solid.
In the intermediate coupling regime, the itinerancy of electrons must be taken into account and there is an antiferromagnetic metal phase between the paramagnetic metal phase and the antiferromagnetic insulator phase due to the competing of the frustration and the interaction.

In summary, by combining the cellular dynamical mean field approximation with the continuous time quantum Monte Carlo method, we investigate the Hubbard model on Shastry-Sutherland lattice at half filling, and obtain phase diagrams with respect to interaction, frustration and temperature. 
Our result shows that in the present of frustration an antiferromagnetic metal phase exists at low temperature between the paramagnetic metal phase and the antiferromagnetic insulator phase.  
When frustration goes beyond the critical value, magnetic orders are suppressed, and Mott transition leads the system to a paramagnetic insulator, which should be a valence bond solid according to previous studies on Heisenberg model on Shastry-Sutherland lattice. 
We hope our study can provide a new perspective for the property of this lattice.

\section*{Methods}
All calculations reported in this work are carried out by using the cellular dynamical mean field theory (CDMFT)\cite{PhysRevLett.87.186401,PhysRevLett.101.186403,RevModPhys.77.1027} and the continuous time quantum Monte Carlo method (CTQMC)\cite{MC, MCRMP}. In our work we map the original lattice onto a four-site effective cluster (see Fig.~1 (b)) embedded in a self-consistent medium. From an initialization of the self-energy $\Sigma(i\omega)$, the effective medium $g(i\omega)$ can be obtained via the coarse-grained Dyson equation,
\begin{equation}
	 g^{-1}(i\omega)=\left[\sum\limits_{k}\frac{1}{i\omega+\mu-t(k)-\Sigma(i\omega)}\right]^{-1}+\Sigma(i\omega),
\end{equation}
where $\mu$ is the chemical potential and $t(k)$ is the Fourier-transformed hopping matrix and $k$ is summed over the reduced Brillouin zone. 

After getting $t(k)$, we can obtain the cluster Green's function $G(i\omega)$ by simulating the effective cluster model using CTQMC as the impurity solver. Using Dyson function $\Sigma=g^{-1}(i\omega)-G^{-1}(i\omega)$ we renew the cluster self-energy $\Sigma(i\omega)$ and complete the iteration.  Here, $g(i\omega)$, $t(k)$, $G(i\omega)$, $\Sigma(i\omega)$ are all $4\times4$ matrices.  We repeat this self-consistent iterative loop until the results are converged, and in each iteration we take $10^7$ Monte Carlo steps.

\clearpage
\newpage

\begin{addendum}
\item [Acknowledgments]
This work was supported by the NKBRSFC under grants Nos. 2011CB921502, 2012CB821305, NSFC under grants Nos. 61227902, 61378017, 11311120053.

\item [Author Contributions]
H. D. L. conceived the idea and designed the research and performed calculations. 
H. D. L., Y. H. C, H. F. L, H. S. T. and W. M. L. contributed to the analysis and interpretation of the results and prepared the manuscript. 

\item [Competing Interests]
The authors declare no competing financial interests.

\item [Correspondence]
Correspondence and requests for materials should be addressed to Hai-Di Liu or Wu-Ming Liu.

\end{addendum}

\clearpage
\newpage
\bigskip
\textbf{Figure 1.}
(a) Two dimensional lattice structure of $Cu^{2+}$ in $SrCu_2(BO_3)_2$; (b) Sketch of Shastry-Sutherland lattice, which is topologically equivalent to (a). The dashed green line marks the four-site cluster which contains four atoms labeled by 1, 2, 3, and 4. $t_{1}$ and $t_2$ are the nearest-neighbor hopping energy and the diagonal hopping energy respectively. We set the distance between the nearest neighbors as the length unit and $t_1=1.0$ as the energy unit. (c) First Brillouin zone of Shastry-Sutherland lattice. $T$, $K$, and $\Gamma$ denote high symmetry points in the first Brillouin zone. 
(d) Tight-binding band structure for $t_2=1.0$; (e) Non-interacting density of states at half-filling for $t_{2}=0.8$, $t_2=1.0$, and $t_{2}=1.2$.

\bigskip
\textbf{Figure 2.}
Non-interacting band dispersion at (a) $t_2=0.8$, (b) $t_2=1.0$, (c) $t_2=1.5$,  (d) $t_2=2.0$, (e) $t_2=2.1$, and (f) $t_2=2.5$. We label I II III and IV from top to bottom. Band II and III begin to separate when $t_2=2.0$, and the system becomes a band insulator at half-filling. We focus on the case when $t_2<2.0$ in this report. 

\bigskip
\textbf{Figure 3.}
(a) Density of state (DOS) for different $U$ with $T=0.2$ and $t_{2}=1.0$. Mott metal-insulator transition happens at $U=9.5$ where an obvious gap appears around the Fermi level. DOS when b) $T=0.1$, $t_2=1.0$, c) $T=0.1$, $t_2=0.8$, and d) $T=0.1$, $t_2=1.2$ are also plotted, where the transition points are $U=8$, $6.5$, and $9.1$ respectively.

\bigskip
\textbf{Figure 4.}
Evolution of double occupancy ($Docc$) as a function of $U$ for $T=0.1$, $0.2$, and $0.5$ when $t_2=1.0$. The blue, red, and black arrows mark the critical $U$'s of Mott transition for $T=0.1$, $0.2$ and $0.5$, and the values are $8$, $9.5$, and $11.8$ respectively. Inset: The evolution of $Docc$ as a function of $t_2$ for different on-site interaction at $T=0.1$.

\bigskip
\textbf{Figure 5.}
Evolution of the staggered magnetic order parameter $m$ and the single-particle gap $\triangle E$  as a function of $U$ at $T=0.1$ and $t_{2}=1.0$. When the interaction is weak, $\triangle E=0$ $m=0$, and the system is in a paramagnetic (PM) metal phase. With the increasing of the interaction $U$, an antiferromagnetic (AFM) metal phase found with $m \not =0$ but $\triangle E=0$. When $U$ is strong enough, $\triangle E \not=0$ $m \not=0$, and the system enters an antiferromagnetic (AFM) insulator phase. The insert picture is for $t_{2}=1.4$.

\bigskip
\textbf{Figure 6.} $t_2-U$ phase diagram at $T=0.1$ is illustrated in (b). Schematic diagrams of a1) paramagnet, a2) antiferromagnet, and a3) dimer phase are also shown. 
When $t_2 < 1.3$, there is a antiferromagnetic metal phase between the paramagnetic metal phase and antiferromagnetic insulator phase. For $t_2>1.3$, a low-temperature paramagnetic insulator phase emerges [which should be a  valence bond state (VBS)].

\bigskip
\textbf{Figure 7.}
 $T-U$ Phase diagram of interacting fermions on the Shastry-Sutherland Lattice at $t_{2}=1.0$. The black line indicates the transition form metal to insulator and the red line shows the transition from paramagnetic (PM) phase to antiferromagnetic (AFM) phase. When the temperature is low enough, with the increasing $U$, there exists a region of antiferromagnetic metal phase before the system enters the antiferromagnetic insulator phase. The insert picture is for the case when $t_{2}=1.3$.

\bigskip
\textbf{Figure 8.}
The distribution of spectral weight at zero frequency for different $U$ at $T=0.1$, with (a) $t_2=1.2$, 
(b) $t_2=1.0$, and (c) $t_2=0.7$.  Peaks in the diagrams represent the dominate distribution of electrons with zero energy in momentum space and thus correspond to the location of Fermi surface. When the effect of interaction is small, it behaves like sharp peaks on the two diagonals in the first Brillouin zone, reflecting the Fermi surface at half-filling.
With the increasing $U$ and decreasing $t_2$, the renormalization effect becomes stronger, and the distribution spread. In (c3) where Mott transition occurs and the system is in the antiferromagnetic phase, no clear patterns of the distribution can be seen.

\clearpage
\newpage

\begin{figure}
  \begin{center}
   \epsfig{file=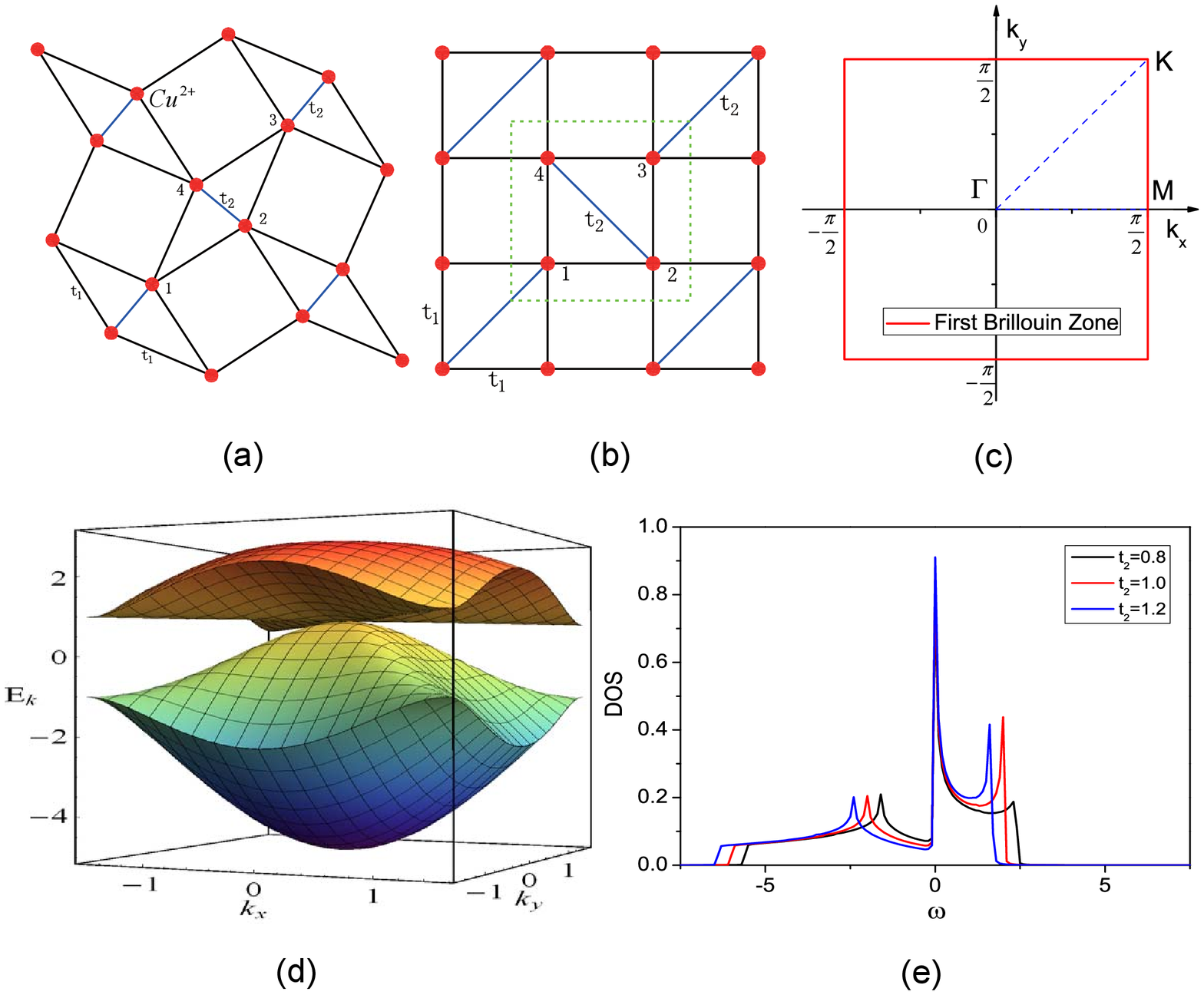,width=14cm}
  \end{center}
   \label{fig1} 
\end{figure}

\newpage
\begin{figure}
  \begin{center}
   \epsfig{file=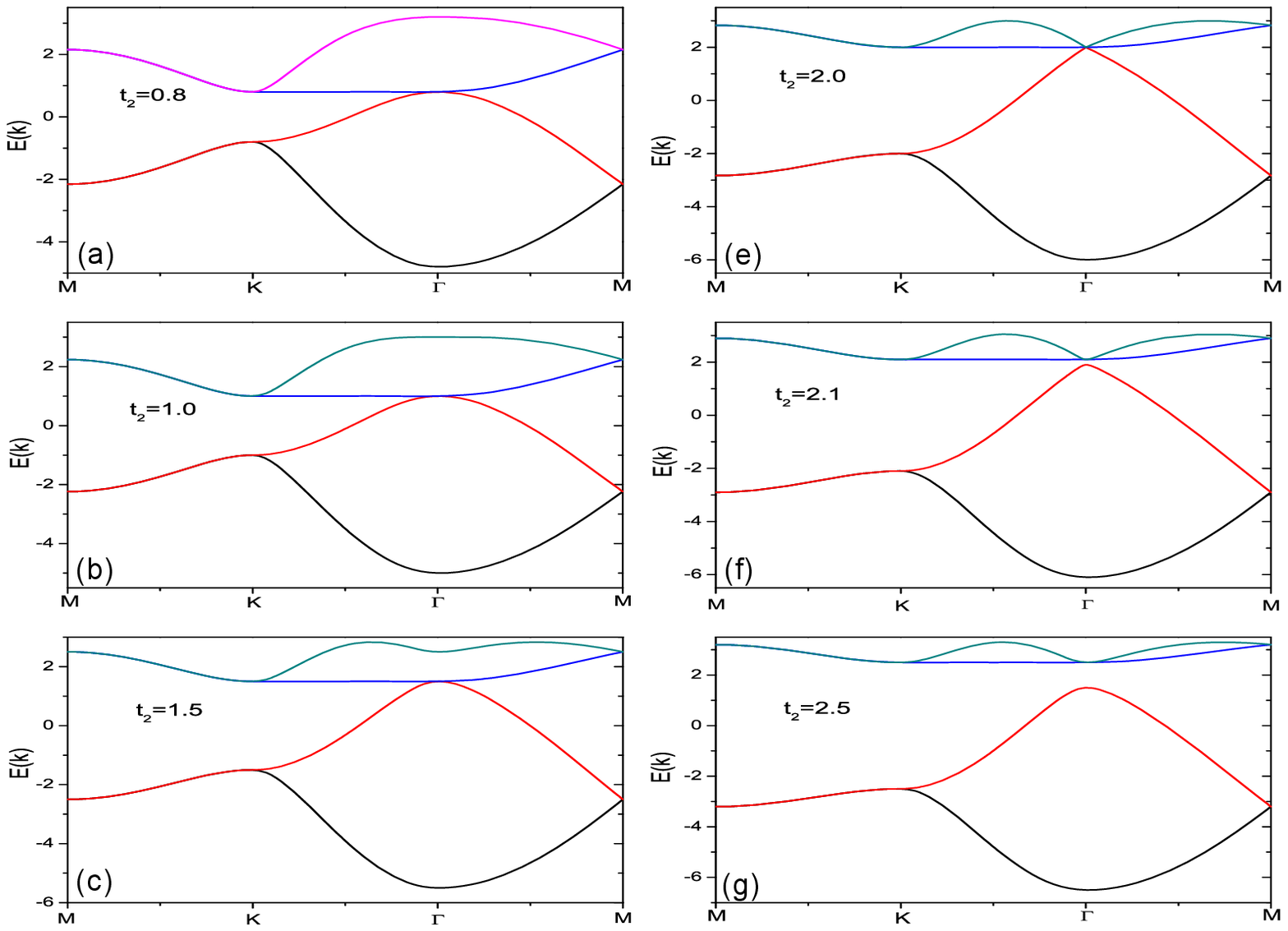,width=14cm}
  \end{center}
 \label{fig2} 
\end{figure}

\newpage
\begin{figure} 
  \begin{center}
   \epsfig{file=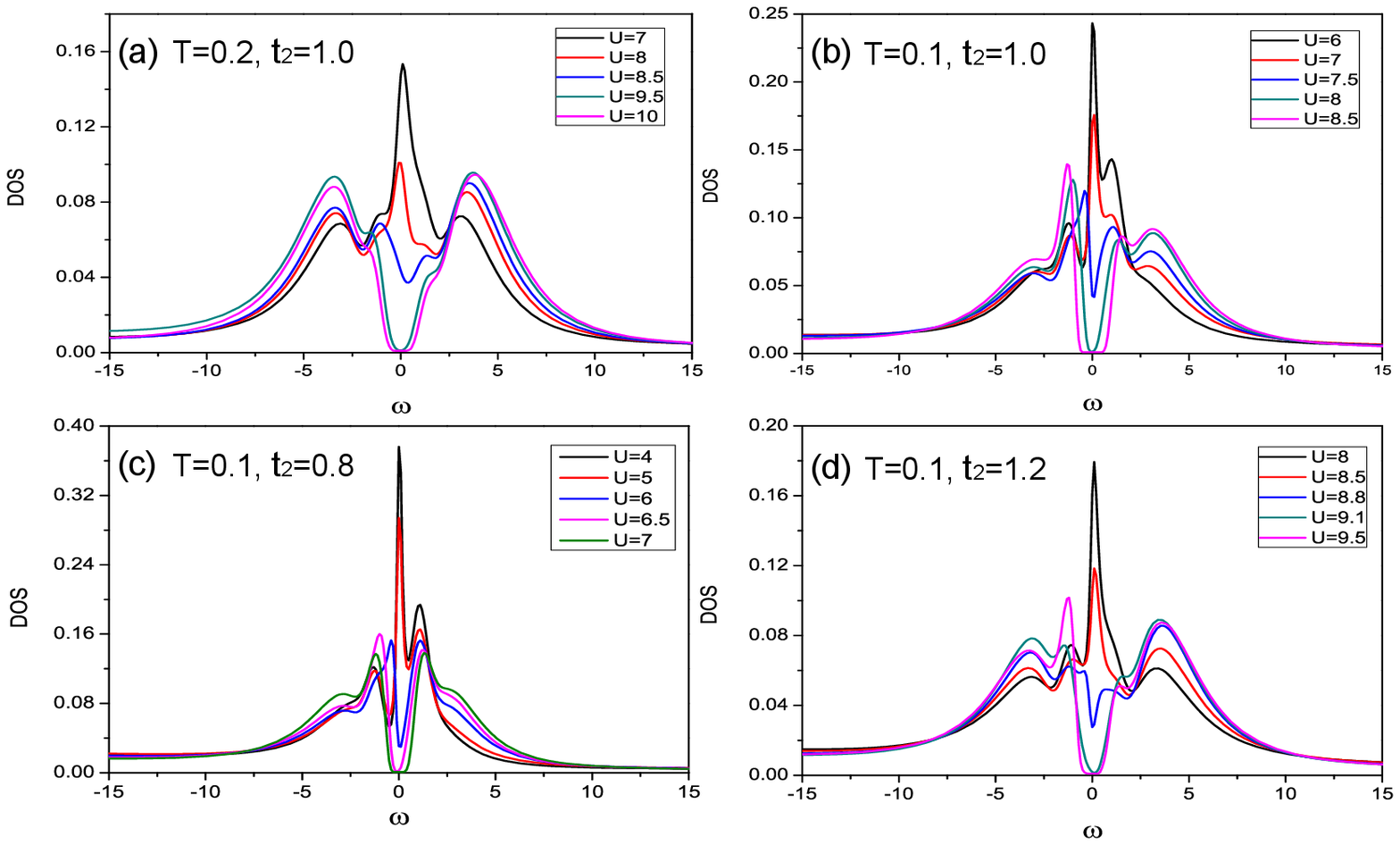,width=14cm}
  \end{center}
 \label{fig3} 
\end{figure}

\newpage

\begin{figure}
   \begin{center}
   \epsfig{file=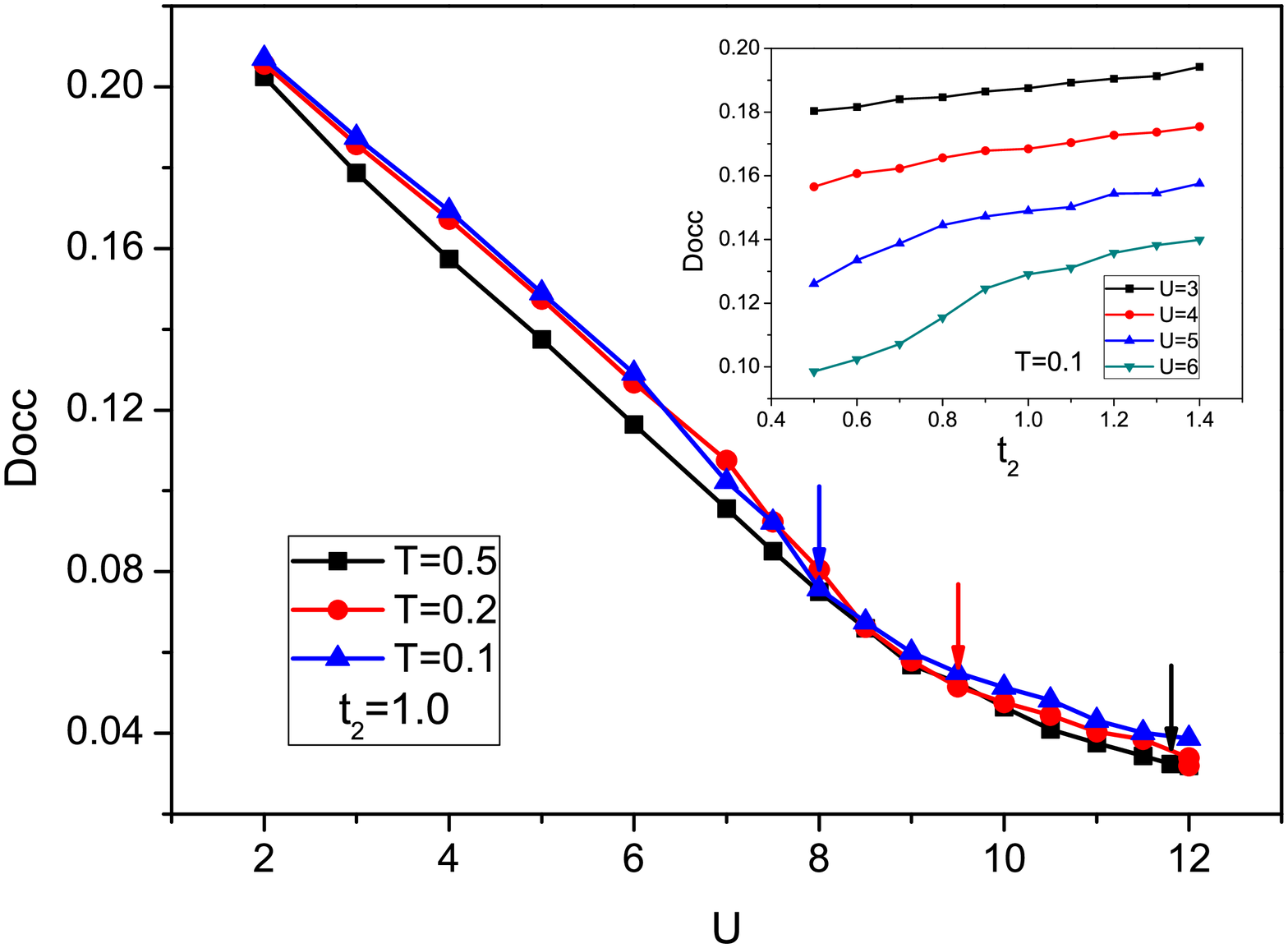,width=14cm}
  \end{center}
 \label{fig4}
\end{figure}

\newpage

\begin{figure}
  \begin{center}
   \epsfig{file=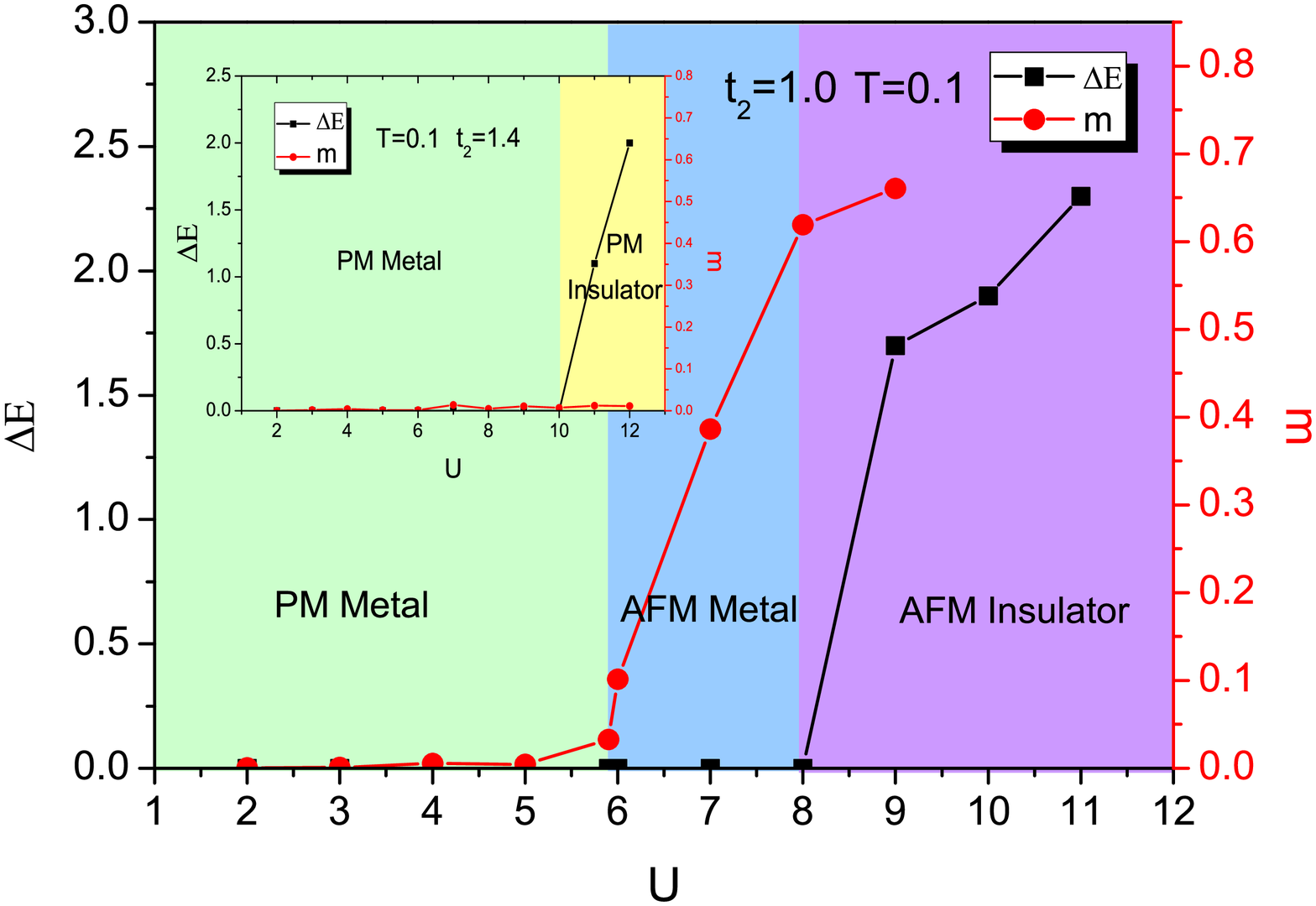,width=14cm}
  \end{center}
 \label{fig5} 
\end{figure}

\newpage

\begin{figure}
  \begin{center}
   \epsfig{file=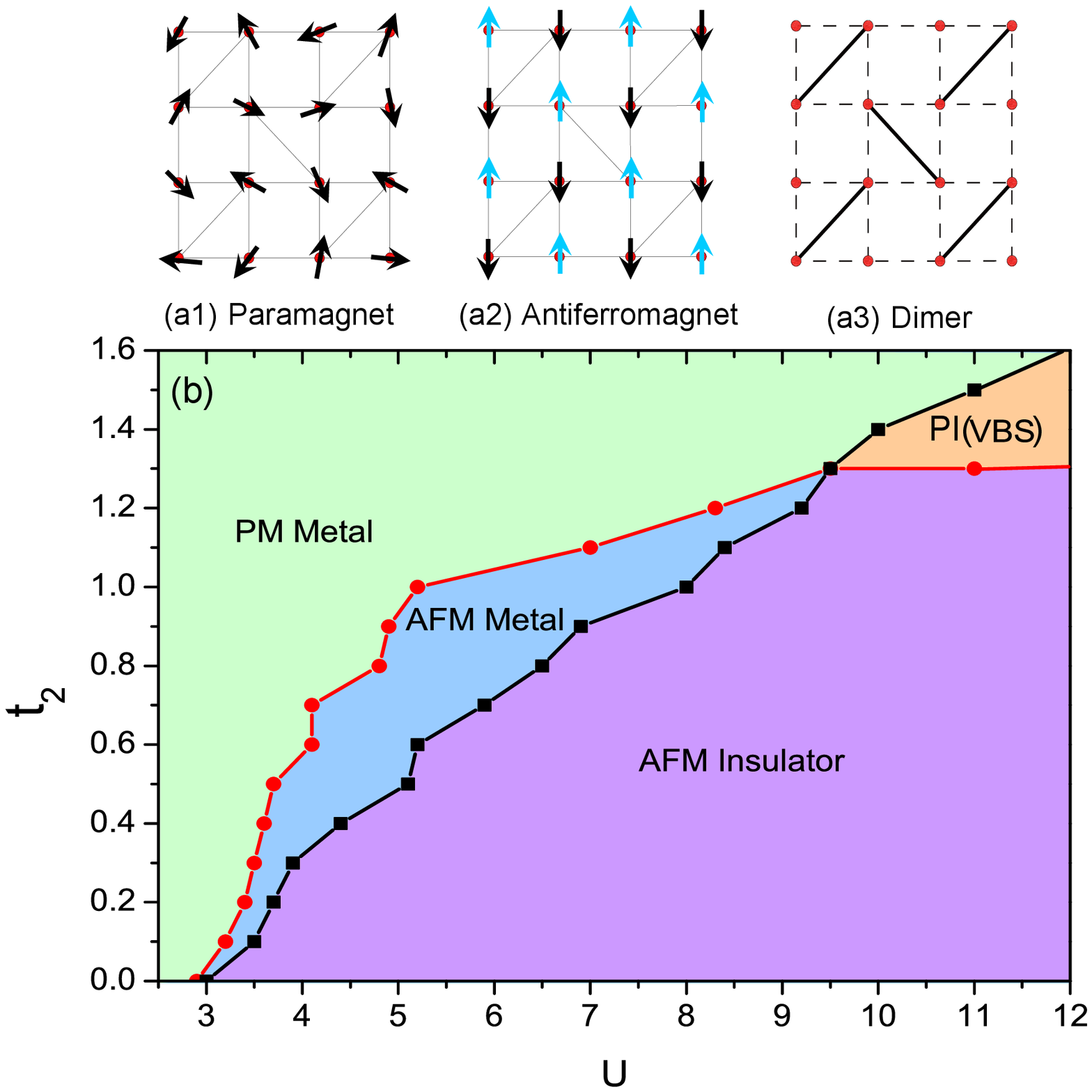,width=14cm}
  \end{center}
 \label{fig6} 
\end{figure}

\newpage

\begin{figure}
  \begin{center}
   \epsfig{file=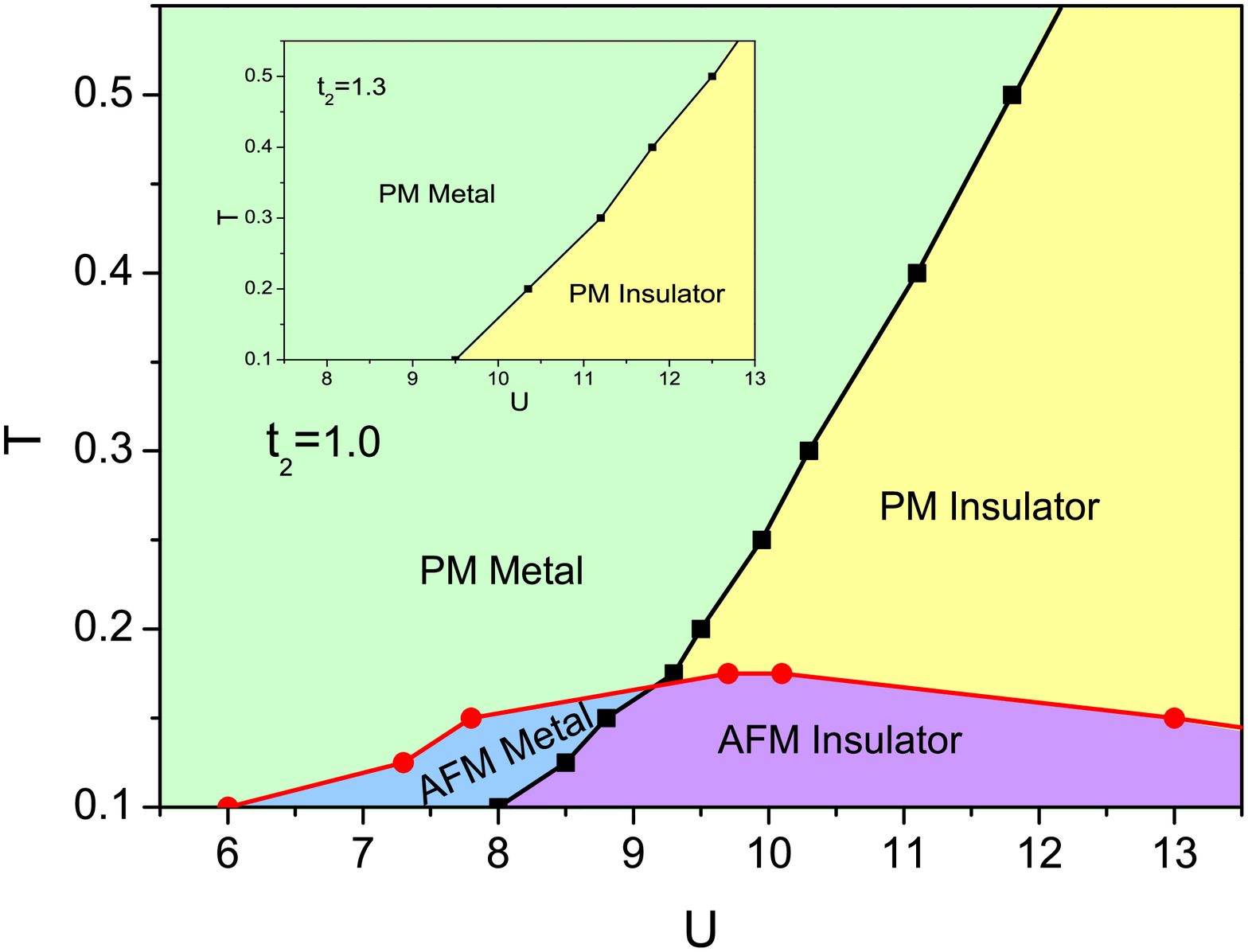,width=14cm}
  \end{center}
 \label{fig7} 
\end{figure}

\newpage

\begin{figure}
  \begin{center}
   \epsfig{file=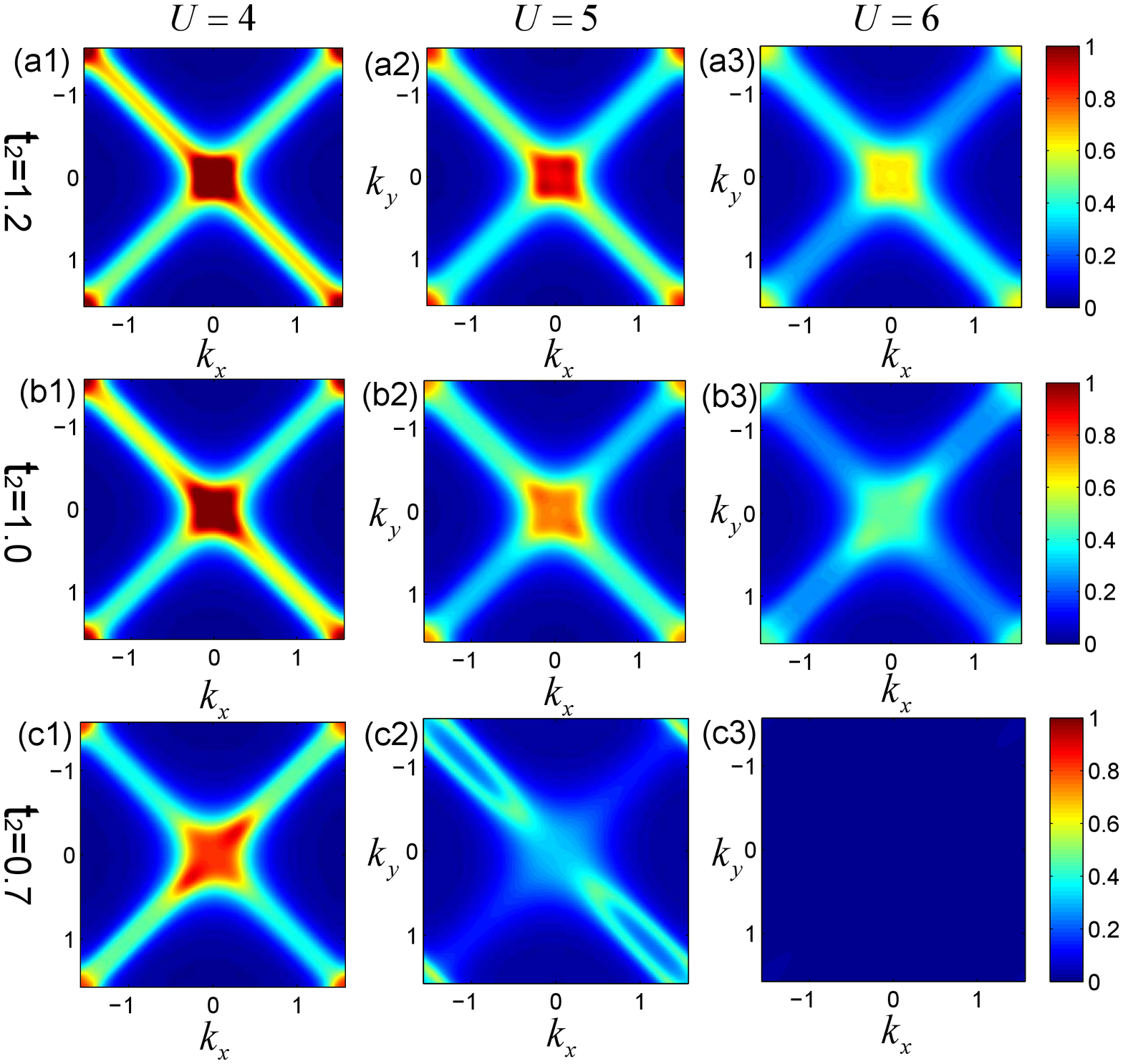,width=14cm}
  \end{center}
 \label{fig8} 
\end{figure}

\end{document}